\documentclass[11pt,a4paper]{article}

\usepackage{jcappub}

\usepackage{color}

\newcommand{\fett}[1]{\boldsymbol{#1}}

\newcommand{\dd}{{\rm{d}}}
\newcommand{\ii}{{\rm{i}}}

\newcommand{\be}{\begin{equation}}
\newcommand{\ee}{\end{equation}}

\definecolor{darkred}{rgb}{0.5,0,0}
\definecolor{darkgreen}{rgb}{0,0.5,0}
\definecolor{darkblue}{rgb}{0,0,0.5}

\usepackage{hyperref}
\hypersetup{ colorlinks,
linkcolor=darkblue,
filecolor=darkgreen,
urlcolor=darkred,
citecolor=darkblue }

\begin{document}

\begin{flushright}
{\large \tt 
TTK-12-15}
\end{flushright}

\title{The recursion relation in Lagrangian perturbation theory}

\date{\today}

\author{Cornelius Rampf}

\affiliation{Institut f\"ur Theoretische Teilchenphysik und Kosmologie, RWTH Aachen, D--52056 Aachen, Germany}

\emailAdd{rampf@physik.rwth-aachen.de}

\abstract{
We derive a recursion relation in the framework of Lagrangian perturbation theory, appropriate for studying the inhomogeneities of the large scale structure of the universe. 
We use the fact that the perturbative expansion
of the matter density contrast is in one-to-one correspondence with standard perturbation theory 
(SPT) at any order. 
This correspondence has been recently shown to be valid up to fourth order for a non-relativistic, irrotational and dust-like component. Assuming it to be valid at arbitrary (higher) order, 
we express the Lagrangian displacement field in terms of the perturbative kernels of SPT, which are itself given by their own and well-known recursion relation. 
We argue that the Lagrangian solution always contains more non-linear information in comparison with the SPT solution, (mainly) if the non-perturbative density contrast is restored after the displacement field is obtained.
}

\maketitle   

\flushbottom
\section{Introduction}

Analytic techniques for studying the inhomogeneities of the large scale structure (LSS) usually rely on several assumptions and approximations: 
({a}) the LSS is formed due to the evolution of gravitational instability only;
({b}) the equations of motion are solved for a pressureless component of cold dark matter particles in terms of a perturbative series on an exactly homogeneous and isotropic background;
({c}) the use of the Newtonian limit, i.e., we demand a non-relativistic fluid, restrict to subhorizon scales, and assume negligible curvature;
({d}) there are no vorticities on sufficient large scales, and also primordial vorticity is absent;
({e}) the use of the single-stream approximation, i.e., neglecting velocity dispersion and higher-order moments of the distribution function; 
({f}) the smoothing volume over the spiky Klimontovich number density is set to zero, i.e., neglecting backreaction effects on the velocity dispersion and on the gravitational field strength. 
We shall consider these restrictions in the current paper. They are appropriate for studying the weakly non-linear regime of structure formation. For departures of this framework see for example  \cite{Durrer:1993db,Ehlers:1996wg,adlerbuchert,Scoccimarro:2000zr,Buchert:2005xj,McDonald:2006hf,Pietroni:2011iz,Buchert:2011yu}.

Perhaps the most straightforward analytic technique is called Eulerian (or standard) perturbation theory (SPT), since the equations are evaluated as a function of Eulerian coordinates \cite{Bernardeau:2001qr}.
Here, the local density contrast $\delta(\fett{x},t) \!\equiv\! [\rho(\fett{x},t) - \overline{\rho}(t)]/ \overline{\rho}(t)$ and the velocity field of the fluid particle are the perturbed quantities. Importantly, the series in SPT relies on the smallness of these fields,
and therefore breaks down as soon as their local values deviate significantly from its mean values.

A convenient way to circumvent this drawback is to use the Lagrangian perturbation theory (LPT) \cite{Buchert:1992ya,Bouchet:1992uh,Bouchet:1994xp,Munshi:1994zb,Sahni:1995rr,Taylor:1996ne,zeldovich:fragmentation,zeldovich:fragmentation2,zeldovich:rev}. In LPT, there is only one perturbed quantity, namely the displacement field $\fett{\Psi}$. It parametrises the gravitationally induced deviation of the particle trajectory field from the homogeneous background
expansion. Therefore, the LPT series does not rely on the smallness of the density and velocity fields, but on the smallness of the deviation of the trajectory field. Furthermore, the explicit extrapolation of the Lagrangian solution leads to improved predictions even in the highly non-linear regime, whereas the series in SPT fails by construction \cite{Zeldovich:1969sb}.

Historically, a great advantage of SPT with respect to LPT was the discovery of a 
simple recursion relation  \cite{Goroff:1986ep}, whereas it was widely believed that there is
no recursion relation in LPT \cite{Bernardeau:2001qr}. 
A common argument for this absence was the additional complicacy in LPT, that Lagrangian transverse fields are also needed to provide an irrotational motion in the Eulerian frame. In this paper we derive an easy expression to maintain the Eulerian irrotationality, thus constraining the Lagrangian 
transverse fields $\fett{\Psi}_T^{(n)}$ at each order $n$. Furthermore, we derive a relation to constrain the Lagrangian longitudinal fields $\fett{\Psi}_L^{(n)}$, such that finally one can construct the displacement field $\fett{\Psi}= \sum_n \fett{\Psi}^{(n)}$ in terms of the aforementioned longitudinal and transverse fields:  $\fett{\Psi}^{(n)}\! =\! \fett{\Psi}_L^{(n)}\!+\!\fett{\Psi}_T^{(n)}$.

The LPT recursion relation is based on the fact that the $n$th order density contrast $\delta^{(n)}$ in SPT and LPT are in one-to-one correspondence with each other (while restricting to the initial position limit) \cite{BuchertRampf:2012}. To obtain the displacement field $\fett{\Psi}^{(n)}$ we shall Taylor expand the Lagrangian mass conservation $\delta = 1/J(\fett{\Psi})-1$,
where $J$ is the Jacobian of the transformation from Eulerian to Lagrangian coordinates. 
Importantly, by performing the Taylor expansion one loses the power of the non-perturbative formula. However, this is just a calculational step to obtain the displacement field---due to the inherent non-linearity in $1/J(\fett{\Psi})$, the final use of the Lagrangian result should be concentrated on the unexpanded $\delta$-relation. This point was first noted by Zel'dovich \cite{Zeldovich:1969sb} who obtained an approximate solution by explicit extrapolation far into the non-linear regime.


\section{Formalism}

\renewcommand{\arraystretch}{1.4}

In the Lagrangian framework the only dynamical variable is the displacement field $\fett{\Psi}$. It maps the fluid element from its initial Lagrangian coordinate $\fett{q}$ to the Eulerian coordinate $\fett{x}$ at cosmic time $t$:
\begin{eqnarray} \label{eq:trafo}
  \fett{x}(\fett{q},t) = \fett{q}+ \fett{\Psi}(\fett{q},t) \,.
\end{eqnarray}
We utilise the Jacobian $J =  \det [ \partial \fett{x} / \partial \fett{q} ]$ to describe mass conservation for our non-relativistic fluid:
\begin{eqnarray}
\label{eq:masscons}
 \dd^3 x  = J (\fett{q},t)\, \dd^3 q \,, \qquad 
 \rho(\fett{x},t)\, \dd^3 x = \overline{\rho}(t)\, \dd^3 q  \,,
\end{eqnarray}
where $\rho(\fett{x},t)$ is the Eulerian density field,  and $\overline\rho(t)$ is
 the mean mass density.  Defining the density contrast $\delta$ as 
\begin{eqnarray}
 \rho(\fett{x},t) = \overline{\rho}(t) \left[1+\delta(\fett{x},t) \right] \,, 
\end{eqnarray}
we can use the mass conservation~(\ref{eq:masscons}) to relate the density 
contrast to the displacement field:
\begin{eqnarray}
 \label{deltaReal}
  \delta(\fett{x},t) = \frac{1}{J(\fett{q},t)}-1  \,.
\end{eqnarray}
In Fourier space this is
\cite{Taylor:1996ne}
\begin{eqnarray}
 \label{deltaNew1}
 \tilde{\delta}(\fett{k},t)  
 =  \int \dd^3 q \, e^{\ii\fett{k}\cdot\fett{q}} \left[ e^{\ii\fett{k}\cdot\fett{\Psi}(\fett{q},t)} -1  \right] ,
\end{eqnarray}
where we have used eqs.\,(\ref{eq:trafo}-\ref{deltaReal}) for the last equality.
We shall use the above equation below to relate the series in LPT to its counterpart in SPT.

In LPT, $\fett{\Psi}$ is expanded by a perturbative series, composed of purely longitudinal perturbations, labeled $\tilde{\fett{\Psi}}_L^{(n)}$, and purely transverse perturbations, denoted 
$\tilde{\fett{\Psi}}_T^{(n)}$. As mentioned before we require an irrotational motion in the Eulerian frame, but Lagrangian transverse fields are mandatory to maintain this constraint.
It is important to note that there is no decoupling between the transverse and longitudinal components. 
Thus, transverse and longitudinal fields depend on \emph{both} transverse and longitudinal fields.
 In Fourier space, the perturbation ansatz is:
\begin{eqnarray} \label{series}
    \tilde{\fett{\Psi}} (\fett{k},t) = \sum_{n=1}^\infty  \tilde{\fett{\Psi}}^{(n)}(\fett{k},t) = \sum_{n=1}^\infty \left[ 
   \tilde{\fett{\Psi}}_L^{(n)}(\fett{k},t)  + \tilde{\fett{\Psi}}_T^{(n)}(\fett{k},t)   \right] \,.
\end{eqnarray}
The solution for the fastest growing mode is
\begin{align}
\label{decompL}   {\tilde{\fett{\Psi}}}_L^{(n)} (\fett{k},t) =& \! -\ii \,D^n(t)\, {\tilde{\fett{L}}}^{(n)} (\fett{k}) \, , \\ 
\label{decompT} {\tilde{\fett{\Psi}}}_T^{(n)} (\fett{k},t) =& \! -\ii \,D^n(t)\; {\tilde{\fett{T}}}^{(n)} (\fett{k})\, ,
\end{align}
where $D = a$ is the linear growth function in an Einstein-de Sitter (EdS) universe normalised to unity at the present time $t_0$, and
\begin{align}
 \label{longitudinal} {\tilde{\fett{L}}}^{(n)} (\fett{k}) &= \int \frac{\dd^3 p_1 \cdots \dd^3 p_n}{(2\pi)^{3n}} 
  \, (2\pi)^3 \delta_D^{(3)}(\fett{p}_{1\cdots n} -\fett{k} ) \, \fett{S}_L^{(n)} (\fett{p}_1, \ldots, \fett{p}_n) 
  \, \tilde{\delta}_0 (\fett{p}_1) \cdots \tilde{\delta}_0 (\fett{p}_n) \,, \\
 \label{transverse} {\tilde{\fett{T}}}^{(n)} (\fett{k}) &= \int \frac{\dd^3 p_1 \cdots \dd^3 p_n}{(2\pi)^{3n}} 
  \, (2\pi)^3 \delta_D^{(3)}(\fett{p}_{1\cdots n} -\fett{k} )  \,
   \fett{S}_T^{(n)} (\fett{p}_1, \ldots, \fett{p}_n) 
  \, \tilde{\delta}_0 (\fett{p}_1) \cdots \tilde{\delta}_0 (\fett{p}_n) \,,
\end{align}
where we have employed the shorthand notation $\fett{p}_{1\cdots n}\!=\! \fett{p}_1 \!+\! \fett{p}_2 \!+\cdots +\! \fett{p}_n$, and $\tilde{\delta}_0 \!\equiv\! \tilde{\delta}^{(1)}(t\!=\!t_0)$ is the linear density contrast;  the vectors $\fett {S}_L^{(n)}$ and  $\fett{S}_T^{(n)}$ 
are the symmetrised longitudinal and transverse kernels respectively, and they reflect the mode-couplings induced by non-linear evolution. The case $n\!=\!1$ denotes the Zel'dovich approximation \cite{Zeldovich:1969sb}, i.e., we restrict to the initial position limit \cite{BuchertRampf:2012}, and in this case the perturbative kernels are simply
$\fett {S}_L^{(1)}(\fett{p})\!=\! \fett{p}/p^2$ and  $\fett{S}_T^{(1)}(\fett{p})\! =\! \fett{0}$, with $|\fett{p}| \!=\! p$. The Zel'dovich solution for the Fourier transform of the displacement field  is thus
\begin{align}
  \label{ZAlong} {\tilde{\fett{\Psi}}}_L^{(1)} (\fett{k},t) =& \! 
     -\ii \,D(t)\, {\tilde{\fett{L}}}^{(1)} (\fett{k}) =   
    -\ii \,D(t)\, {\fett{S}}_L^{(1)} (\fett{k}) \, \tilde{\delta}_0 (\fett{k})
   =  -\ii \,D(t)\, \frac{\fett{k}}{k^2} \tilde{\delta}_0 (\fett{k}) \, , \\
  \label{ZAtrans} {\tilde{\fett{\Psi}}}_T^{(1)} (\fett{k},t) =& \, \fett{0} \,,
\end{align}
where we have used eqs.~(\ref{decompL})-(\ref{transverse}).
In general, we expect longitudinal vectors of the form $\fett {S}_L^{(n)}\!=\!\fett{p}_{12 \cdots n}/p_{12 \cdots n}^2 B_L^{(n)}$, where $B_L^{(n)}$ is a scalar function to be determined, and the (non-zero) transverse kernels are constrained by $\fett{p}_{12 \cdots n} \cdot \fett {S}_T^{(n)}\! =\! 0$.  
In the following, we describe how to obtain solutions for the longitudinal
 and transverse displacements, $\tilde{\fett{\Psi}}_L^{(n)}$ and $\tilde{\fett{\Psi}}_T^{(n)}$ for $n\geq 2$.

\section{From SPT to LPT}\label{sec:density}

In SPT, the density contrast is a perturbed quantity, and the solution for an EdS universe is:
\begin{eqnarray}
 \tilde{\delta}(\fett{k},t) =  \sum_{n=1}^\infty  D^n(t) \, \tilde{\delta}^{(n)}(\fett{k}) \,,
\end{eqnarray}
with
\begin{align}
  {\tilde{{\delta}}}^{(n)} (\fett{k}) &=  \int \frac{\dd^3 p_1 \cdots \dd^3 p_n}{(2\pi)^{3n}} 
  \, (2\pi)^3 \delta_D^{(3)}(\fett{p}_{1\cdots n} -\fett{k} ) F_n^{(s)} (\fett{p}_1, \ldots, \fett{p}_n) 
  \, \tilde{\delta}_0 (\fett{p}_1) \cdots \tilde{\delta}_0 (\fett{p}_n) \, ,
\end{align}
where the symmetrised kernels $F_n^{(s)}$ are given by the well-known SPT recursion relation \cite{Goroff:1986ep,Jain:1993jh,Bernardeau:2001qr}. But note the special case $F_1^{(s)} =1$.
Now, using the above results for the displacement field and the density contrast, we Taylor expand 
eq.\,(\ref{deltaNew1}) and collect all individual $n$th order terms in eq.\,(\ref{deltaNew1}).  The $n$th order density contrast reads then
\begin{align}
 \label{deltaLag1}
  \tilde{\delta}^{(n)}(\fett{k}) &= \int \frac{\dd^3 p_1 \cdots \dd^3 p_n}{(2\pi)^{3n}}   \, (2\pi)^3 \delta_D^{(3)}(\fett{p}_{1\cdots n} - \fett{k}) \, X_n^{(s)}(\fett{p}_1, \ldots, \fett{p}_n) \, \tilde{\delta}_0 (\fett{p}_1) \cdots \tilde{\delta}_0 (\fett{p}_n) \,,
\end{align}
where we have defined the symmetric scalars $X_n^{(s)}$, which can be found in \cite{BuchertRampf:2012} up to fourth order.
 Explicitly, for $n\!= \! 1,2$ and $3$ they are
\begin{align}
 \label{x1} {X}_{1}^{(s)}(\fett{p}_1) &=  \fett{k} \cdot \fett{S}_{L \oplus T}^{(1)}(\fett{p}_1)  \,, \\
 \label{x2} {X}_{2}^{(s)}(\fett{p}_1,\fett{p}_2)  &=  \fett{k}\cdot \fett{S}_{L \oplus T}^{(2)}(\fett{p}_1,\fett{p}_2)   +\frac 1 2 \fett{k} \cdot \fett{S}_{L \oplus T}^{(1)}(\fett{p}_1) \,\fett{k} \cdot \fett{S}_{L \oplus T}^{(1)}(\fett{p}_2) \,, \\
 \label{thirdX} {X}_{3}^{(s)}(\fett{p}_1,\fett{p}_2,\fett{p}_3) &= \fett{k} \cdot 
 \fett{S}_{L \oplus T}^{(3)}(\fett{p}_1,\fett{p}_2,\fett{p}_3) 
  + \frac 1 6 \fett{k} \cdot \fett{S}_{L \oplus T}^{(1)}(\fett{p}_1) \,\fett{k} \cdot \fett{S}_{L \oplus T}^{(1)}(\fett{p}_2) \,\fett{k} \cdot \fett{S}_{L \oplus T}^{(1)}(\fett{p}_3)  \nonumber \\
   & \qquad+  \frac 1 3 \left\{ \fett{k} \cdot \fett{S}_{L \oplus T}^{(1)}(\fett{p}_1)\, \fett{k}\cdot \fett{S}_{L \oplus T}^{(2)}(\fett{p}_2,\fett{p}_3) + \text{two perms.}  \right\} \,,  
\end{align}
where we have introduced the short-hand notation $\fett{S}_{L \oplus T}^{(n)} \equiv \fett{S}_{L}^{(n)}+\fett{S}_{T}^{(n)}$.
The Dirac-delta in eq.\,(\ref{deltaLag1}) fixes $\fett{k} = \fett{p}_{1\cdots n}$ and as an important note we recognise the equivalence of the density contrast in both SPT and LPT:
\begin{eqnarray}
 \label{assump}
  X_n^{(s)}(\fett{p}_1,\ldots ,\fett{p}_n)\Big|_{\fett{k} = \fett{p}_{1\cdots n}} \Big. = F_n^{(s)}(\fett{p}_1,\ldots ,\fett{p}_n) \,.
\end{eqnarray}
Above relation has been proven to be valid at least up to the fourth order \cite{BuchertRampf:2012}.
Physically, this result is not surprising, because the transformation to Lagrangian coordinates should not alter the density contrast (neither the velocity of the fluid particle changes). It is thus a reliable assumption to demand eq.~(\ref{assump}) to arbitrary order, which we do so in the following.

In our setting, the requirement of an irrotational fluid motion in Eulerian space implies vanishing (Lagrangian) transverse contributions up to the second order, i.e., $\fett{S}_{L \oplus T}^{(1)} \equiv \fett{S}_{L}^{(1)}$ and $\fett{S}_{L \oplus T}^{(2)} \equiv \fett{S}_{L}^{(2)}$. We will discuss this issue in the following section.
This means that we can immediately obtain the longitudinal solution  $\fett{k} \cdot \fett{S}_{L}^{(2)}$ by equating eq.\,(\ref{x2}) and $F_2^{(s)}$ with the constraint $\fett{k} \!=\! \fett{p}_{12}$ and the use of the Zel'dovich approximation, i.e., eqs.~(\ref{ZAlong}) and~(\ref{ZAtrans}):
\begin{align}
 \label{bl2} \fett{p}_{12} \cdot \fett{S}_{L}^{(2)}(\fett{p}_1,\fett{p}_2) = F_2^{(s)}(\fett{p}_1,\fett{p}_2) - \frac 1 2 \left( 1+\frac{\fett{p}_1 \cdot \fett{p}_2}{p_1^2} \right) \left( 1+\frac{\fett{p}_1 \cdot \fett{p}_2}{p_2^2} \right) =: B_L^{(2)}(\fett{p}_1,\fett{p}_2)  \,,
\end{align}
and as the $\fett{S}_L^{(n)}$'s originate from longitudinal perturbations (i.e., the field perturbations can be written in terms of a potential), we have
\begin{align}
 \fett{S}_{L}^{(2)} (\fett{p}_1,\fett{p}_2) = \frac{\fett{p}_{12}}{p_{12}^2} B_L^{(2)}(\fett{p}_1,\fett{p}_2) \,,
\end{align}
and the second order longitudinal displacement field is then
\begin{align}
  \tilde{\fett{\Psi}}_L^{(2)}(\fett{k},t) &=  -\ii D^2(t)  {\tilde{\fett{L}}}^{(2)} (\fett{k}) \,,
\intertext{with}
  {\tilde{\fett{L}}}^{(2)} (\fett{k}) &= \int \frac{\dd^3 p_1 \dd^3 p_2}{(2\pi)^{6}} 
  \, (2\pi)^3 \delta_D^{(3)}(\fett{p}_{12} -\fett{k} )\,  \frac{\fett{p}_{12}}{p_{12}^2} B_L^{(2)}(\fett{p}_1,\fett{p}_2) 
  \, \tilde{\delta}_0 (\fett{p}_1) \tilde{\delta}_0 (\fett{p}_2) \,.
\end{align}
 This simple procedure works as long as the displacement field is fully longitudinal in the Lagrangian frame. However, since $\fett{p}_{12} \cdot  \fett{S}_{T}^{(2)}(\fett{p}_1,\fett{p}_2)\! =\!0$ due to the transverseness condition, eq.\,(\ref{x2}) gives us the longitudinal solution $\fett{p}_{12} \cdot \fett{S}_{L}^{(2)}(\fett{p}_1,\fett{p}_2)$ only. Thus, even if there \emph{were} a transverse part already at second order, it would not affect the second order density contrast. This leads to an important consequence for the density contrast at arbitrary order $n$: the transverse kernel $\fett{S}_{T}^{(m)}$ cannot be obtained via the relation (\ref{assump}), but it will affect the $n$th order density contrast $\tilde\delta^{(n)}$ if $m\!<\!n$. Thus, $\fett{S}_{T}^{(m)}$ (or $\tilde{\fett{T}}^{(m)}$) has to be constrained at each order.

\section{Lagrangian transverse fields}\label{sec:trans}

The Eulerian irrotationality condition states the vanishing of the Eulerian curl of the particle motion \cite{Buchert:1987xy,Bernardeau:2001qr}:
\be \label{irrot}
 \fett{\nabla}_{\fett{x}} \times \fett{u}(\fett{x},t) = \fett{0} \,.
\ee
 As mentioned in the introduction this implies a restriction to sufficient large scales, where we expect that the time evolution should not generate vorticities. Because of the non-trivial transformation to Lagrangian coordinates, however, the displacement field must include a transverse component as well \cite{Buchert:1992ya}. Crucially, longitudinal and transverse fields are dynamically coupled in the Lagrangian picture,\footnote{In SPT the velocity field can be written in 
longitudinal and transverse fields, and they are decoupled from each other.} even if the transverse fields are zero---the latter is the case at lower orders. At higher orders, transverse fields attach the same importance as longitudinal fields.

 It is possible to write the irrotationality condition directly in Fourier space, and it leads to the general result for $n \geq 2$:
\begin{align} \label{FTtrans}
 \tilde{\fett{T}}^{(n)}(\fett{k})  &=  \int \frac{\dd^3 p_1 \dd^3 p_2}{\left(2\pi \right)^6} \left(2\pi\right)^3 \delta_D^{(3)}(\fett{p}_{12} -\fett{k} ) \left[ \fett{p}_1 \left( \fett{p}_2 \cdot \fett{p}_{12} \right) -\fett{p}_2 \left( \fett{p}_1 \cdot \fett{p}_{12} \right)  \right]/p_{12}^2  \nonumber \\
&\qquad\hspace{0.208cm}\times \sum_{ \substack{1 \leq i \leq j,\\ i+j=n}}  \Big\{ \Big. \frac{j-i}{n} \!\!  \Big. \left[ \tilde{\fett{L}}^{(i)}(\fett{p}_1) + \tilde{\fett{T}}^{(i)}(\fett{p}_1) \right] \cdot \left[  \tilde{\fett{L}}^{(j)}(\fett{p}_2) + \tilde{\fett{T}}^{(j)}(\fett{p}_2) \right]  \Big\} \,,
\intertext{or, alternatively}
\tilde{\fett{T}}^{(n)}(\fett{k}) &=  \frac{\fett{k}}{k^2} \times \int \frac{\dd^3 p_1 \dd^3 p_2}{\left(2\pi \right)^6} \left(2\pi\right)^3 \delta_D^{(3)}(\fett{p}_{12} -\fett{k} )  
  \,\, \left(\fett{p}_1 \times \fett{p}_2 \right)  \nonumber \\
&\qquad\hspace{0.208cm}\times
 \sum_{ \substack{1 \leq i \leq j,\\ i+j=n}}  \Big\{ \Big. \frac{j-i}{n} \!\!  \Big. \left[ \tilde{\fett{L}}^{(i)}(\fett{p}_1) + \tilde{\fett{T}}^{(i)}(\fett{p}_1) \right] \cdot \left[
  \tilde{\fett{L}}^{(j)}(\fett{p}_2) + \tilde{\fett{T}}^{(j)}(\fett{p}_2) \right]
  \Big\} \,.
\end{align}
This representation of the original irrotationality condition~(\ref{irrot}) is new, since it is
 exact at any order in perturbation theory and it embeds transverse sources as well;\footnote{Reference \cite{Bernardeau:2008ss} gives a Fourier expression for the irrotationality condition as well, however it is non-exact.} additionally, it contains a technical simplification with respect to the commonly used irrotationality condition which we shall highlight in the appendix \ref{app}. With the above expression, the transverse displacement field with time evolution is then
$\tilde{\fett{\Psi}}_T^{(n)}(\fett{k},t)\!=\! -\ii\,D^n(t) \, \tilde{\fett{T}}^{(n)}(\fett{k})$.
The perturbation vectors in the curly brackets are given by eqs.\,(\ref{longitudinal}) and~(\ref{transverse}), thus the explicit Lagrangian formalism is not needed and only the (lower order) results have to be plugged in.

With the use of the Zel'dovich approximation, we immediately obtain for $n\!=\!2$: $\tilde{\fett{T}}^{(2)} \!=\! \fett{0}$ and thus $\fett{S}_T^{(2)}\!=\!\fett{0}$, because of the trivial condition $j-i\!=\!0$ but also due to symmetry reasons in the integrations over $\fett{p}_1$ and $\fett{p}_2$.  In general, there are vanishing contributions for $i\!=\!j$ in the summation.
Note that in a very general treatment, it is possible to obtain non-vanishing transverse solutions already at the second order \cite{Buchert:1993xz}. We are not considering this case here.

\section{The recursive procedure}

Assuming that eq.\,(\ref{assump}) is valid for arbitrary order $n$, we obtain the $n$th order longitudinal displacement field  iteratively in terms of $F_n^{(s)}$ 
and lower order LPT results. Projecting out the longitudinal part, i.e.,
$
  B_L^{(n)}(\fett{p}_1,\ldots, \fett{p}_n) \! \equiv\! \fett{p}_{1\cdots n} \cdot \fett{S}_L^{(n)}(\fett{p}_1,\ldots, \fett{p}_n) \,,
$
we obtain for the longitudinal part of the displacement field $\fett{\Psi}^{(n)}$:
\begin{eqnarray}
 \label{rec1}
 \fett{S}_L^{(n)} = \frac{\fett{p}_{1\cdots n}}{p_{1\cdots n}^2} B_L^{(n)} \,, \qquad B_L^{(n)} = F_n^{(s)} - E_n^{(s)} \,, 
\end{eqnarray}
with
\begin{align}
\label{rec2} E_n^{(s)} \equiv {\cal O}_n^{(s)}\left\{ \sum_{a=1}^{n}  \frac{\left[ \fett{p}_{1\cdots n} \cdot \left( \fett{S}_{L \oplus T}^{(1)}+\cdots +\fett{S}_{L \oplus T}^{(n-1\geq 2)} \right) \right]^{a}}{a!}  \right\} \,. 
\end{align}
This is our second main result. The last line is the strict consequence of Taylor expanding eq.~(\ref{deltaNew1}), and we have introduced 
the operator ${\cal O}_n^{(s)}\{X \}$ which extracts the symmetric $n$th order part of its argument $X$, e.g. 
\begin{align}
 \label{recDens}
{\cal O}_3^{(s)}\left\{\fett{k}\cdot \fett{S}^{(1)}(\fett{p}_1) \right. \left.+ \fett{k}\cdot \fett{S}^{(1)}(\fett{p}_1) \,\fett{k} \cdot\fett{S}^{(2)}(\fett{p}_2,\fett{p}_3)\right\} = 
  &\frac 1 3 \Big\{ \Big. \fett{k} \cdot \fett{S}^{(1)}(\fett{p}_1) \,\fett{k} \cdot \fett{S}^{(2)}(\fett{p}_2,\fett{p}_3) \nonumber \\ &\quad \hspace{2.5cm} + \rm{two\;perms.} \Big. \Big\}\,.
\end{align}
The dependence of the vectors in eq.\,(\ref{rec2}) is
$\fett{S}_{L \oplus T}^{(k)}\! \equiv\! \fett{S}_{L \oplus T}^{(k)}(\fett{p}_1,\ldots, \fett{p}_k)$, for products of two vectors it 
is $\fett{p}_{1\cdots k}\!\cdot\! \fett{S}_{L \oplus T}^{(i)} \,\fett{p}_{1\cdots k} \!\cdot\! \fett{S}_{L \oplus T}^{(j)}\! \equiv \!\fett{p}_{1\cdots k}\!\cdot\! \fett{S}_{L \oplus T}^{(i)}(\fett{p}_1,\ldots, \fett{p}_i) \,\fett{p}_{1\cdots k} 
\!\cdot\! \fett{S}_{L \oplus T}^{(j)}(\fett{p}_{j+1},\ldots, \fett{p}_j)$, with 
$k\! =\! i+j$, and similar for higher order products. 

It is then straightforward to calculate higher order displacement fields.

\section{Example: The third order displacement field}\label{examp}

In this section we demonstrate how the third order displacement field can be obtained and expressed in terms of the SPT kernels $F_n^{(s)}$.
We start with eqs.\,(\ref{rec1}) and~(\ref{rec2}) for $n\!=\!3$, this leads to the longitudinal solution
\be 
  \fett{S}_L^{(3)} = \frac{\fett{p}_{123}}{p_{123}^2} B_L^{(3)}\,, \qquad B_L^{(3)}(\fett{p}_1,\fett{p}_2,\fett{p}_3 ) = F_3^{(s)}(\fett{p}_1,\fett{p}_2,\fett{p}_3 ) - E_3^{(s)}(\fett{p}_1,\fett{p}_2,\fett{p}_3 ) \,, 
\ee
with
\begin{align}
  E_3^{(s)}(\fett{p}_1,\fett{p}_2,\fett{p}_3 ) &=  
   \frac 1 6 \left( 1+\frac{\fett{p}_1 \cdot \fett{p}_{23}}{p_1^2}  \right)  \left( 1+\frac{\fett{p}_2 \cdot \fett{p}_{13}}{p_2^2}  \right)  \left( 1+\frac{\fett{p}_3 \cdot \fett{p}_{12}}{p_3^2}  \right)  \nonumber \\
   & \quad+  \frac 1 3 \left\{  \left( 1+\frac{\fett{p}_1 \cdot \fett{p}_{23}}{p_1^2}  \right) \, 
  \left(1 + \frac{\fett{p}_1 \cdot \fett{p}_{23}}{p_{23}^2} \right) B_L^{(2)}(\fett{p}_2,\fett{p}_3) + \text{two perms.}  \right\} \,, 
\end{align}
where $B_L^{(2)} = F_2^{(s)} - E_2^{(s)}$ is given in eq.\,(\ref{bl2}). To obtain  the longitudinal displacement field 
at third order we use eqs.~(\ref{decompL}) and~(\ref{longitudinal}). This leads to
\begin{align}
  \tilde{\fett{\Psi}}_L^{(3)}(\fett{k},t) &=  -\ii D^3(t) \int \frac{\dd^3 p_1 \dd^3 p_2 \dd^3 p_3}{(2\pi)^{9}} 
  \, (2\pi)^3 \delta_D^{(3)}(\fett{p}_{123} -\fett{k} )\,  \tilde{\delta}_0 (\fett{p}_1) \,\tilde{\delta}_0 (\fett{p}_2)\, \tilde{\delta}_0 (\fett{p}_3)  \nonumber \\
   & \qquad \hspace{6.8cm}\, \times \frac{\fett{p}_{123}}{p_{123}^2} B_L^{(3)}(\fett{p}_1,\fett{p}_2,\fett{p}_3)  \,.
\end{align}
On the other hand, the transverse field~(\ref{FTtrans}) at third order is
\begin{align}
 \tilde{\fett{T}}^{(3)}(\fett{k}) =   &\int \frac{\dd^3 p_1 \dd^3 p_2}{\left(2\pi \right)^6} \left(2\pi\right)^3 \delta_D^{(3)}(\fett{p}_{12} -\fett{k} ) \frac{\fett{k}}{3k^2} \times 
 \left( \fett{p}_1 \times \fett{p}_2 \right)  \, 
   \tilde{\fett{L}}^{(1)}(\fett{p}_1) \cdot  \tilde{\fett{L}}^{(2)}(\fett{p}_2) \,. 
\end{align}
The only thing we have to do is to use the lower order results $\tilde{\fett{L}}^{(1)}$ and $\tilde{\fett{L}}^{(2)}$ and substitute the integration limits in the above expression. We then have for the transverse displacement field~(\ref{decompT}) at third order
\begin{align}
 \tilde{\fett{\Psi}}_T^{(3)}(\fett{k},t) =  &-\ii D^3(t) \int \frac{\dd^3 p_1 \dd^3 p_2 \dd^3 p_3}{\left(2\pi \right)^9} \left(2\pi\right)^3 \delta_D^{(3)}(\fett{p}_{123} -\fett{k} ) \, 
   \frac{\fett{k}}{3k^2} \times \left( \fett{p}_1 \times \fett{p}_{23}  \right)
 \nonumber \\ &\quad  \hspace{3.5cm}\times  
   B_L^{(2)}(\fett{p}_2,\fett{p}_3) \, \frac{\fett{p}_1 \cdot \fett{p}_{23}}{p_1^2 p_{23}^2} \, \tilde{\delta}_0 (\fett{p}_1)  \, \tilde{\delta}_0 (\fett{p}_2) \,  \tilde{\delta}_0 (\fett{p}_3)  \,,
\end{align}
and thus we obtain
\be
   \tilde{\fett{\Psi}}^{(3)}(\fett{k},t) =  \tilde{\fett{\Psi}}_L^{(3)}(\fett{k},t)+  \tilde{\fett{\Psi}}_T^{(3)}(\fett{k},t) \,.
\ee
In general, the use of the recursion relation reduces the work significantly. The final expressions for higher order displacement fields are surely longer, but the procedure is exactly the same compared to the above.

\section{Conclusions}

For the first time, we have formulated an iterative procedure to calculate the Fourier transform of the Lagrangian displacement field up to arbitrary order in perturbation theory. 
Our procedure is based on the physical assumption that the density contrast agrees in SPT and LPT, if the treatment is perturbative and if we restrict our formalism to the initial position limit (IPL) (in the IPL, the linear density contrast is evaluated in the vicinity of the initial Lagrangian position instead of the evolved Eulerian coordinate;
see the thorough discussions in \cite{BuchertRampf:2012,RampfWong:2012}). This allows us to relate the LPT series to its counterpart in SPT through the density contrast relation~(\ref{deltaNew1}), and the SPT results are given by the well known SPT recursion relation. 

Even for an irrotational motion in the Eulerian frame,
the Lagrangian displacement field consists not only of longitudinal components but 
also of transverse components; the transverse perturbations affect the $n$th order density contrast if they are of lower order than $n$. As a consequence, the transverse perturbations cannot be calculated within the density contrast relation, but have to be constrained at each order. We have 
calculated a new representation of the irrotationality condition directly in Fourier space, eq.\,(\ref{FTtrans}). This new representation has the big advantage that the explicit Lagrangian
formalism is not needed and only the (lower order) results have to be plugged in.
The calculation of the Lagrangian transverse fields is straightforward, and so is then the full (i.e., longitudinal and transverse) displacement field at any order.

Some remarks for applications are appropriate here. First of all, the Lagrangian solution always contains more non-linear information than the standard one due to the inherent non-linearity of the unexpanded relation of the density contrast~(\ref{deltaReal}). In a future project we shall introduce a numerical treatment of the very non-perturbative expression, and we will clarify the performance of higher order LPT solutions.
Furthermore, the use of our result is not restricted to the IPL: One may relax this approximation after the iterative procedure, thus effectively readjusting the inherent level of non-linearities (the kernels derived in the IPL are still valid).
Finally, (higher order) LPT solutions are for example needed for resummation techniques of matter poly-spectra (e.g.~\cite{Matsubara:2007wj,Okamura:2011nu}).

\begin{acknowledgments}
CR would like to thank Y.~Y.~Y.~Wong, T.~Buchert, V.~Zheligovsky, U.~Frisch and the referee for valuable comments on the manuscript.
\end{acknowledgments}

\appendix

\section{The derivation of equation~(\ref{FTtrans})}\label{app}

The Eulerian irrotationality constraint is, ($\fett{x}$ is the Eulerian coordinate)
\be
  \fett{\nabla_x} \times \fett{u}(\eta,\fett{x}) = \fett{0} \,,
\ee
where $\fett{u}$ is the peculiar velocity flow, and we use the superconformal time $\dd \eta = \dd t /a^2$, with $a \propto t^{2/3} \propto 1/\eta^2$ for an Einstein-de Sitter universe. We set the initial vorticity to zero.
In Lagrangian space, the requirement of an irrotational fluid motion yields \cite{BuchertRampf:2012}:
\vskip-0.3cm \begin{align}
 \label{LN4fin} \varepsilon_{ijk} \frac{\dd}{\dd \eta} \Psi_{k,j} &- \varepsilon_{ijk}  \Psi_{l,j} \frac{\dd}{\dd \eta} \Psi_{l,k} =   \Psi_{i,n} \,\varepsilon_{njk} \left( \Psi_{l,j} \frac{\dd}{\dd \eta} \Psi_{l,k} - \frac{\dd}{\dd \eta} \Psi_{k,j} \right) \,, 
\end{align}
with $\fett{\Psi} \!\equiv \!\fett{x}-\fett{q}$ being the non-perturbative displacement field.
Summation over repeated indices is implied, and the subscript ',$j$' denotes a partial derivative with respect to the Lagrangian coordinate $q_j$.  As before we decompose $\fett{\Psi}$ in a longitudinal
and transverse part which we denote by $\fett{\Psi}_L$ and $\fett{\Psi}_T$, respectively, and similar for their $n$th order parts: $\fett{\Psi}^{(n)} = \fett{\Psi}_L^{(n)}+\fett{\Psi}_T^{(n)}$.
Note that $\fett{\nabla_q} \times \fett{\Psi}_L^{(n)} \!= \!\fett{0}$ because of $\fett{\Psi}_L^{(n)}\! \equiv\! \fett{\nabla_q} \phi^{(n)}$ and due to the symmetry of the second derivatives.
Equation (\ref{LN4fin}) is the strict result of the transformation into Lagrangian coordinates. However,
the term in brackets on the RHS is essentially redundant, since it is always of higher order. The reason for this can be understood, if we rewrite the above equation in a schematic but perturbative way: 
\begin{align}
C_i^{(n)} &= - \sum_{p+q=n} \Psi_{i,m}^{(p)} C_m^{(q)} \,,
\intertext{where we have defined}
 C_i^{(n)} &\equiv \sum_{p+q=n} \varepsilon_{ijk} \left( \frac{\dd}{\dd \eta} \Psi_{k,j}^{(n)} 
  - \Psi_{l,j}^{(p)} \frac{\dd}{\dd \eta} \Psi_{l,k}^{(q)} \right) \,.
\end{align} The leading order solution is therefore at any order just
\be
C_i^{(n)}\! =  \sum_{p+q=n} \varepsilon_{ijk} \left( \frac{\dd}{\dd \eta} \Psi_{k,j}^{(n)} 
  - \Psi_{l,j}^{(p)} \frac{\dd}{\dd \eta} \Psi_{l,k}^{(q)} \right) \equiv 0 \,,
\ee
 The $n$th order solution of eq.\,(\ref{LN4fin}) is thus
\vskip-0.5cm \begin{align} \label{TransPert}
 &\varepsilon_{ijk} \frac{\dd}{\dd \eta} \left(\fett{\Psi}_T^{(n)} \right)_{k,j} =  \sum_{p+q=n}
\varepsilon_{ijk} \left( \fett{\Psi}_L^{(p)}+\fett{\Psi}_T^{(p)} \right)_{l,j} \frac{\dd}{\dd \eta} \left( \fett{\Psi}_L^{(q)}+\fett{\Psi}_T^{(q)} \right)_{l,k} \,.
\end{align} 
Furthermore, by denoting that the time evolution of the $n$th order displacement is $\propto \eta^{-2n}$ ($\equiv D$), we can separate the time evolution of the $n$th order displacement from its longitudinal and transverse part: $\fett{\Psi}_L^{(n)} (\eta,\fett{q})\equiv \fett{L}^{(n)}(\fett{q}) \,\eta^{-2n}$, and 
 $\fett{\Psi}_T^{(n)}(\eta,\fett{q}) \equiv \fett{T}^{(n)}(\fett{q}) \,\eta^{-2n}$. Then, we can evaluate the temporal derivatives in Eq.\,(\ref{TransPert}) and  obtain
\be \label{mainTrans}
  \varepsilon_{ijk} {T}_{k,j}^{(n)} = \frac 1 2 \sum_{0<p<n} \frac{n-2p}{n}  \varepsilon_{ijk} \left(\fett{L}+\fett{T}\right)_{l,j}^{(p)} 
  \left( \fett{L}+\fett{T}\right)_{l,k}^{(n-p)} \,.
\ee
This is our final result of the Eulerian irrotationality condition in real space. Note that the above is now a quadratic equation and not a cubic 
one anymore (cf.~our starting point, Eq.\,(\ref{LN4fin})). Equation\,(\ref{mainTrans})  contains therefore a dramatic simplification, which holds as long as we demand the series approximation and have vanishing initial vorticity.

To obtain the counterpart of Eq.\,(\ref{mainTrans}) in Fourier space, only minor manipulations are necessary:  
Since $\fett{T}^{(n)}$ is purely transverse, we can write it in terms of a vector potential: $T_k^{(n)} \equiv \varepsilon_{klm} A_{m,l}^{(n)}$. 
Then, we Fourier transform the above equation and multiply it with an additional Levi-Civita connection. This cancels out the gauge ambiguity which resulted from the introduction of the vector potential $\fett{A}$. These considerations then yield Eq.\,(\ref{FTtrans}).



\end{document}